\documentclass[twoside,twocolumn]{article}

\usepackage{mathtools}
\usepackage{amsmath,amssymb}
\usepackage{booktabs}
\usepackage{natbib}

\usepackage[T1]{fontenc} 
\usepackage{microtype} 

\usepackage[english]{babel} 

\usepackage[hmarginratio=1:1,top=32mm,columnsep=20pt]{geometry} 
\usepackage[hang, small,up,up]{caption} 
\usepackage{booktabs} 

\usepackage{lettrine} 

\usepackage{enumitem} 
\setlist[itemize]{noitemsep} 

\usepackage{abstract} 

\usepackage{titlesec} 
\renewcommand\thesection{\Roman{section}} 
\renewcommand\thesubsection{\roman{subsection}} 
\titleformat{\section}[block]{\large\scshape\centering}{\thesection.}{1em}{} 
\titleformat{\subsection}[block]{\large}{\thesubsection.}{1em}{} 

\usepackage{fancyhdr} 
\usepackage{titling} 

\usepackage{hyperref} 

\pretitle{\begin{center}\Huge\bfseries} 
\posttitle{\end{center}} 
\title{Learning to Rank Scientific Documents from the Crowd} 
\author{%
\textsc{Jesse M Lingeman}\thanks{Corresponding author} \\[1ex] 
\normalsize University of Massachusetts Amherst \\ 
\normalsize \href{mailto:lingeman@cs.umass.edu}{lingeman@cs.umass.edu} 
\and 
\textsc{Hong Yu} \\[1ex] 
\normalsize University of Massachusetts Medical School \\ 
\normalsize \href{mailto:hong.yu@umassmed.edu}{hong.yu@umassmed.edu} 
}
\date{} 
\renewcommand{\maketitlehookd}{%
\begin{abstract}
\noindent \textbf{Motivation}:
Finding related published articles is an important task in any science, but with the explosion of new work in the biomedical domain it has become especially challenging. Most existing methodologies use text similarity metrics to identify whether two articles are related or not. However biomedical knowledge discovery is hypothesis-driven. The most related articles may not be ones with the highest text similarities. In this study, we first develop an innovative crowd-sourcing approach to build an expert-annotated document-ranking corpus. Using this corpus as the gold standard, we then evaluate the approaches of using text similarity to rank the relatedness of articles. Finally, we develop and evaluate a new supervised model to automatically rank related scientific articles. \\
\textbf{Results}:
Our results show that authors' ranking differ significantly from rankings by text-similarity-based models. 
By training a learning-to-rank model on a subset of the annotated corpus, we found the best supervised learning-to-rank model (SVM\textsuperscript{Rank}) significantly surpassed state-of-the-art baseline systems.\\
\textbf{Availability}: Code and data are both publicly available.  \thanks{http://github.com/umassbionlp/crowd-ranking.git}\\
\end{abstract}
}

\DeclarePairedDelimiterX{\norm}[1]{\lVert}{\rVert}{#1}

\begin{document}

\maketitle






\section{Introduction}

The number of biomedical research papers published has increased dramatically in recent years.
As of October, 2016, PubMed houses over 26 million citations, with almost 1 million from the first 3 quarters of 2016 alone \footnote{Accessed on October 6, 2016.}.
It has become impossible for any one person to actually read all of the work being published.
We require tools to help us determine which research articles would be most informative and related to a particular question or document.
For example, a common task when reading articles is to find articles that are most related to another.
Major research search engines offer such a ``related articles'' feature.
However, we propose that instead of measuring relatedness by text-similarity measures, we build a model that is able to infer relatedness from the authors' judgments.

\cite{Wilbur:1994bx} consider two kinds of queries important to bibliographic information retrieval: the first is a search query written by the user and the second is a request for documents most similar to a document already judged relevant by the user.
Such a query-by-document (or query-by-example) system has been implemented in the \textit{de facto} scientific search engine PubMed---called Related Citation Search.
\cite{Lin:2007bn} show that 19\% of all PubMed searches performed by users have at least one click on a related article. 
Google Scholar provides a similar Related Articles system.
Outside of bibliographic retrieval, query-by-document systems are commonly used for patent retrieval, Internet search, and plagiarism detection, amongst others.
Most work in the area of query-by-document uses text-based similarity measures (\cite{Ganguly:2011tu,Mahdabi:2012gd,KIM:2014ik}).
However, scientific research is hypothesis driven and therefore we question whether text-based similarity alone is the best model for bibliographic retrieval.
In this study we asked authors to rank documents by ``closeness'' to their work.
The definition of ``closeness'' was left for the authors to interpret, as the goal is to model which documents the authors subjectively feel are closest to their own. 
Throughout the paper we will use ``closeness'' and ``relatedness'' interchangeably.

We found that researchers' ranking by closeness differs significantly from the ranking provided by a traditional IR system.
Our contributions are three fold:
\begin{enumerate}
\item	We show that asking biomedical researchers to rank documents related to their work is an easy and effective way to generate customized annotation data.
\item	We show that ranking by a researcher's definition of closeness differs from text-similarity-based ranking.
\item	We implement a supervised learning-to-rank system to better emulate the author's choices, compared to the unsupervised systems used in the past.
\end{enumerate}

The principal ranking algorithms of query-by-document in bibliographic information retrieval rely mainly on text similarity measures (\cite{Lin:2007bn,Wilbur:1994bx}).
For example, the foundational work of \cite{Wilbur:1994bx} introduced the concept of a ``document neighborhood'' in which they pre-compute a text-similarity based distance between each pair of documents.
When a user issues a query, first an initial set of related documents is retrieved.
Then, the neighbors of each of those documents is retrieved, i.e., documents with the highest text similarity to those in the initial set.
In a later work, \cite{Lin:2007bn} develop the PMRA algorithm for PubMed related article search.
PMRA is an unsupervised probabilistic topic model that is trained to model ``relatedness'' between documents.
\cite{Smucker:2006dp} introduce the competing algorithm Find-Similar for this task, treating the full text of documents as a query and selecting related documents from the results.

Outside bibliographic IR, prior work in query-by-document includes patent retrieval (\cite{Xue:2009ir,Mahdabi:2012gd}), finding related documents given a manuscript (\cite{Lin:2007bn,Ontrup:2003vs}), and web page search (\cite{Weng:2011fh,Lee:2012gr}).
Much of the work focuses on generating shorter queries from the lengthy document.
For example, noun-phrase extraction has been used for extracting short, descriptive phrases from the original lengthy text (\cite{Yang:2009cx}).
Topic models have been used to distill a document into a set of topics used to form query (\cite{Nallapati:2008dq}).
\cite{Xue:2009ir} generated queries using the top TF*IDF weighted terms in each document.
\cite{KIM:2014ik} suggested extracting phrasal concepts from a document, which are then used to generate queries.
\cite{Ganguly:2011tu} combined query extraction and pseudo-relevance feedback for patent retrieval.
\cite{Lee:2012gr} employ supervised machine learning model (i.e., Conditional Random Fields) (\cite{Lafferty:2001ty}) for query generation.
\cite{Gobeill:2009ve} explored ontology to identify chemical concepts for queries.

There are also many biomedical-document specific search engines available.
Many information retrieval systems focus on question answering systems such as those developed for the TREC Genomics Track (\cite{Hersh:2009biba}) or BioASQ Question-Answer (\cite{Krithara:2016bp}) competitions.
Systems designed for question-answering use a combination of natural language processing techniques to identify biomedical entities, and then information retrieval systems to extract relevant answers to questions.
Systems like those detailed in \cite{Yang:2015dq} can provide answers to yes/no biomedical questions with high precision.
However what we propose differs from these systems in a fundamental way: given a specific document, suggest the most important documents that are related to it.

The body of work most related to ours is that of citation recommendation.
The goal of citation recommendation is to suggest a small number of publications that can be used as high quality references for a particular article (\cite{Ren:2014ek,Lin:2007bn}).
Topic models have been used to rank articles based on the similarity of latent topic distribution (\cite{Nallapati:2008dq,Tang:2009if,Lin:2007bn}).
These models attempt to decompose a document into a few important keywords.
Specifically, these models attempt to find a latent vector representation of a document that has a much smaller dimensionality than the document itself and compare the reduced dimension vectors.

Citation networks have also been explored for ranking articles by importance, i.e., authority (\cite{Brin:1998jv,Wu:2012hp}).
\cite{Ren:2014ek} introduced heterogeneous network models, called meta-path based models, to incorporate venues (the conference where a paper is published) and content (the term which links two articles, for citation recommendation).
Another highly relevant work is \cite{Weng:2011fh} who decomposed a document to represent it with a compact vector, which is then used to measure the similarity with other documents.
Note that we exclude the work of context-aware recommendation, which analyze each citation's local context, which is typically short and does not represent a full document.

One of the key contributions of our study is an innovative approach for automatically generating a query-by-document gold standard.
Crowd-sourcing has generated large databases, including Wikipedia and Freebase.
Recently, \cite{Ipeirotis:2014ds} concluded that unpaid participants performed better than paid participants for question answering.
They attribute this to unpaid participants being more intrinsically motivated than the paid test takers: they performed the task for fun and already had knowledge about the subject being tested.
In contrast, another study, \cite{Kobren:2014ti}, compared unpaid workers found through Google Adwords (GA) to paid workers found through Amazon Mechanical Turk (AMT).
They found that the paid participants from AMT outperform the unpaid ones.
This is attributed to the paid workers being more willing to look up information they didn't know.
In the bibliographic domain, authors of scientific publications have contributed annotations (\cite{Yu:2010ch}).
They found that authors are more willing to annotate their own publications (\cite{Yu:2010ch}) than to annotate other publications (\cite{Ramesh:2015cj}) even though they are paid.
In this work, our annotated dataset was created by the unpaid authors of the articles.

\section*{Materials and Methods}
\subsection*{Benchmark Datasets}
In order to develop and evaluate ranking algorithms we need a benchmark dataset.
However, to the best of our knowledge, we know of no openly available benchmark dataset for bibliographic query-by-document systems.
We therefore created such a benchmark dataset.

The creation of any benchmark dataset is a daunting labor-intensive task, and in particular, challenging in the scientific domain because one must master the technical jargon of a scientific article, and such experts are not easy to find when using traditional crowd-sourcing technologies (e.g., AMT).
For our task, the ideal annotator for each of our articles are the authors themselves.
The authors of a publication typically have a clear knowledge of the references they cite and their scientific importance to their publication, and therefore may be excellent judges for ranking the reference articles.

Given the full text of a scientific publication, we want to rank its citations according to the author's judgments.
We collected recent publications from the open-access PLoS journals and asked the authors to rank by closeness five citations we selected from their paper.
PLoS articles were selected because its journals cover a wide array of topics and the full text articles are available in XML format.
We selected the most recent publications as previous work in crowd-sourcing annotation shows that authors' willingness to participate in an unpaid annotation task declines with the age of publication (\cite{Yu:2010ch}).
We then extracted the abstract, citations, full text, authors, and corresponding author email address from each document.
The titles and abstracts of the citations were retrieved from PubMed, and the cosine similarity between the PLoS abstract and the citation's abstract was calculated.
We selected the top five most similar abstracts using TF*IDF weighted cosine similarity, shuffled their order, and emailed them to the corresponding author for annotation.
We believe that ranking five articles (rather than the entire collection of the references) is a more manageable task for an author compared to asking them to rank all references.
Because the documents to be annotated were selected based on text similarity, they also represent a challenging baseline for models based on text-similarity features.
In total 416 authors were contacted, and 92 responded (22\% response rate).
Two responses were removed from the dataset for incomplete annotation.

We asked authors to rank documents by how ``close to your work'' they were.
The definition of closeness was left to the discretion of the author.
The dataset is composed of 90 annotated documents with 5 citations each ranked 1 to 5, where 1 is least relevant and 5 is most relevant for a total of 450 annotated citations.

\subsection*{Learning to Rank}
\begin{figure}[!b]
  \centering
  \includegraphics[scale=0.5]{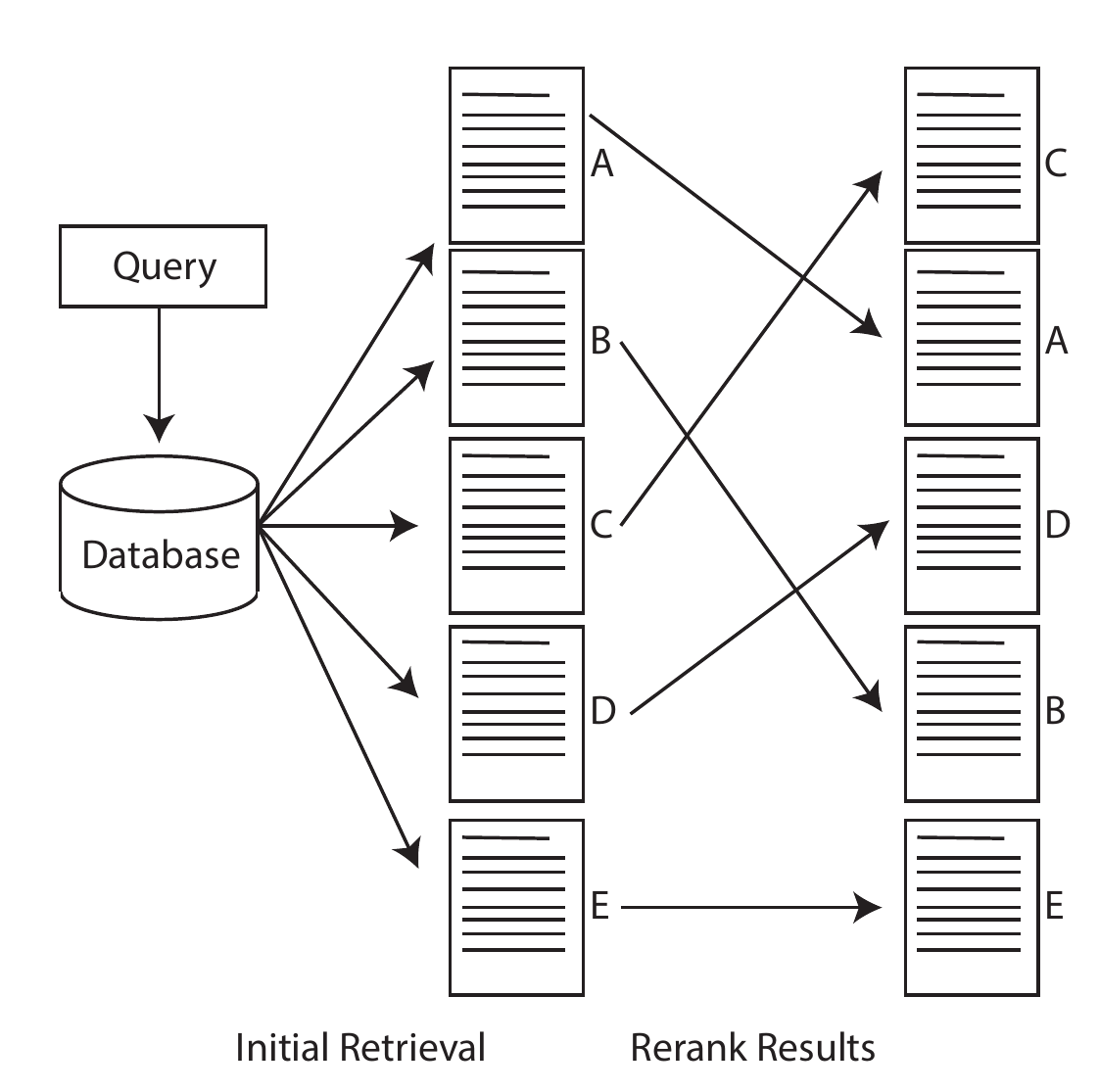}
  \caption{
    The basic pipeline of a learning-to-rank system.
    An initial set of results for a query is retrieved from a search engine, and then that subset is reranked.
    During the reranking phase new features may be extracted.
  }
  \label{l2rfig}
\end{figure}

Learning-to-rank is a technique for reordering the results returned from a search engine query.
Generally, the initial query to a search engine is concerned more with recall than precision: the goal is to obtain a subset of potentially related documents from the corpus.
Then, given this set of potentially related documents, learning-to-rank algorithms reorder the documents such that the most relevant documents appear at the top of the list.
This process is illustrated in Figure \ref{l2rfig}.

There are three basic types of learning-to-rank algorithms: point-wise, pair-wise, and list-wise.
Point-wise algorithms assign a score to each retrieved document and rank them by their scores.
Pair-wise algorithms turn learning-to-rank into a binary classification problem, obtaining a ranking by comparing each individual pair of documents.
List-wise algorithms try to optimize an evaluation parameter over all queries in the dataset.

Support Vector Machine (SVM) (\cite{Cortes:1995p2757}) is a commonly used supervised classification algorithm that has shown good performance over a range of tasks.
SVM can be thought of as a binary linear classifier where the goal is to maximize the size of the gap between the class-separating line and the points on either side of the line.
This helps avoid over-fitting on the training data.
SVM\textsuperscript{Rank} is a modification to SVM that assigns scores to each data point and allows the results to be ranked (\cite{Joachims:2006iz}).
We use SVM\textsuperscript{Rank} in the experiments below.
SVM\textsuperscript{Rank} has previously been used in the task of document retrieval in (\cite{Cao:2006kt}) for a more traditional short query task and has been shown to be a top-performing system for ranking.

SVM\textsuperscript{Rank} is a point-wise learning-to-rank algorithm that returns scores for each document.
We rank the documents by these scores.
It is possible that sometimes two documents will have the same score, resulting in a tie.
In this case, we give both documents the same rank, and then leave a gap in the ranking.
For example, if documents 2 and 3 are tied, their ranked list will be [5, 3, 3, 2, 1].

Models are trained by randomly splitting the dataset into 70\% training data and 30\% test data.
We apply a random sub-sampling approach where the dataset is randomly split, trained, and tested 100 times due to the relatively small size of the data.
A model is learned for each split and a ranking is produced for each annotated document.

We test three different supervised models.
The first supervised model uses only text similarity features, the second model uses all of the features, and the third model runs forward feature selection to select the best performing combination of features.
We also test using two different models trained on two different datasets: one trained using the gold standard annotations, and another trained using the judgments based on text similarity that were used to select the citations to give to the authors.

We tested several different learning to rank algorithms for this work.
We found in preliminary testing that SVM\textsuperscript{Rank} had the best performance, so it will be used in the following experiments.

\subsection*{Features}
Each citation is turned into a feature vector representing the relationship between the published article and the citation.
Four types of features are used: text similarity, citation count and location, age of the citation, and the number of times the citation has appeared in the literature (citation impact).
Text similarity features measure the similarity of the words used in different parts of the document.
In this work, we calculate the similarity between a document $D$ and a document it cites $C$ by transforming the their text into term vectors.
For example, to calculate the similarity of the abstracts between $D$ and $C$ we transform the abstracts into two term vectors, $D_A$ and $C_A$.
The length of each of the term vectors is $|D_A \cup C_A|$.
We then weight each word by its Term-frequency * Inverse-document frequency (TF*IDF) weight.
TF*IDF is a technique to give higher weight to words that appear frequently in a document but infrequently in the corpus.
Term frequency is simply the number of times that a word $w$ appears in a document.
Inverse-document frequency is the logarithmically-scaled fraction of documents in the corpus in which the word $w$ appears.
Or, more specifically:
\begin{equation*}
  \mathrm{idf}(t,D) = \log \dfrac{N}{1 + |\{d \in D : t \in d\}|}
\end{equation*}
where $N$ is the total number of documents in the corpus, and the denominator is the number of documents in which a term $t$ appears in the corpus $D$.
Then, TF*IDF is defined as:
\begin{equation*}
  \mathrm{tfidf}(t,d,D) = \mathrm{tf}(t,d) \mathrm{idf}(t,D)
\end{equation*}
where $t$ is a term, $d$ is the document, and $D$ is the corpus.
For example, the word ``the'' may appear often in a document, but because it also appears in almost every document in the corpus it is not useful for calculating similarity, thus it receives a very low weight.
However, a word such as ``neurogenesis'' may appear often in a document, but does not appear frequently in the corpus, and so it receives a high weight.
The similarity between term vectors is then calculated using cosine similarity:
\begin{equation*}
  \textrm{similarity} = \textrm{cos}(\theta) = \dfrac{\mathbf{A} \cdot \mathbf{B}}{\norm{\mathbf{A}} \norm{\mathbf{B}}}
\end{equation*}
where $A$ and $B$ are two term vectors.
The cosine similarity is a measure of the angle between the two vectors.
The smaller the angle between the two vectors, i.e., the more similar they are, then the closer the value is to 1.
Conversely, the more dissimilar the vectors, the closer the cosine similarity is to 0.

We calculate the text similarity between several different sections of the document $D$ and the document it cites $C$.
From the citing article $D$, we use the title, full text, abstract, the combined discussion/conclusion sections, and the 10 words on either side of the place in the document where the actual citation occurs.
From the document it cites $C$ we only use the title and the abstract due to limited availability of the full text.
In this work we combine the discussion and conclusion sections of each document because some documents have only a conclusion section, others have only a discussion, and some have both.
The similarity between each of these sections from the two documents is calculated and used as features in the model.

The age of the citation may be relevant to its importance.
As a citation ages, we hypothesize that it is more likely to become a ``foundational'' citation rather than one that directly influenced the development of the article.
Therefore more recent citations may be more likely relevant to the article.
Similarly, ``citation impact'', that is, the number of times a citation has appeared in the literature (as measured by Google Scholar) may be an indicator of whether or not an article is foundational rather than directly related.
We hypothesize that the fewer times an article is cited in the literature, the more impact it had on the article at hand.

We also keep track of the number of times a citation is mentioned in both the full text and discussion/conclusion sections.
We hypothesize that if a citation is mentioned multiple times, it is more important than citations that are mentioned only once.
Further, citations that appear in the discussion/conclusion sections are more likely to be crucial to understanding the results.
We normalize the counts of the citations by the total number of citations in that section.
In total we select 15 features, shown in Table \ref{feature-results-table}.
The features are normalized within each document so that each of citation features is on a scale from 0 to 1, and are evenly distributed within that range.
This is done because some of the features (such as years since citation) are unbounded.

\subsection*{Baseline Systems}
We compare our system to a variety of baselines.
(1) Rank by the number of times a citation is mentioned in the document.
(2) Rank by the number of times the citation is cited in the literature (citation impact).
(3) Rank using Google Scholar Related Articles.
(4) Rank by the TF*IDF weighted cosine similarity.
(5) Rank using a learning-to-rank model trained on text similarity rankings.
The first two baseline systems are models where the values are ordered from highest to lowest to generate the ranking.
The idea behind them is that the number of times a citation is mentioned in an article, or the citation impact may already be good indicators of their closeness.
The text similarity model is trained using the same features and methods used by the annotation model, but trained using text similarity rankings instead of the author's judgments.

We also compare our rankings to those found on the popular scientific article search engine Google Scholar.
Google Scholar is a ``black box'' IR system: they do not release details about which features they are using and how they judge relevance of documents.
Google Scholar provides a ``Related Articles'' feature for each document in its index that shows the top 100 related documents for each article.
To compare our rankings, we search through these related documents and record the ranking at which each of the citations we selected appeared.
We scale these rankings such that the lowest ranked article from Google Scholar has the highest relevance ranking in our set.
If the cited document does not appear in the set, we set its relevance-ranking equal to one below the lowest relevance ranking found.

Four comparisons are performed with the Google Scholar data.
(1) We first train a model using our gold standard and see if we can predict Google Scholar's ranking.
(2) We compare to a baseline of using Google Scholar's rankings to train and compare with their own rankings using our feature set.
(3) Then we train a model using Google Scholar's rankings and try to predict our gold standard.
(4) We compare it to the model trained on our gold standard to predict our gold standard.

\subsection*{Evaluation Measures}

\paragraph{NDCG}
Normalized Discounted Cumulative Gain (NDCG) is a common measure for comparing a list of estimated document relevance judgments with a list of known judgments (\cite{Croft:2009ws}).
To calculate NDCG we first calculate a ranking's Discounted Cumulative Gain (DCG) as:
\begin{equation}
  \text{DCG} = \text{rel}_1 + \sum\limits_{i=1}^N \dfrac{2^{\text{rel}_i} - 1}{\log_2(i+1)}
\end{equation}
where rel$_i$ is the relevance judgment at position $i$.
Intuitively, DCG penalizes retrieval of documents that are not relevant (rel$_i = 0$).
However, DCG is an unbounded value.
In order to compare the DCG between two models, we must normalize it.
To do this, we use the ideal DCG (IDCG), i.e., the maximum possible DCG given the relevance judgments.
The maximum possible DCG occurs when the relevance judgments are in the correct order.
\begin{equation}
  \text{NDCG} = \dfrac{\text{DCG}}{\text{IDCG}}
\end{equation}
The NDCG value is in the range of 0 to 1, where 0 means that no relevant documents were retrieved, and 1 means that the relevant documents were retrieved and in the correct order of their relevance judgments.

\paragraph{Kendall's $\tau$}
Kendall's $\tau$ is a measure of the correlation between two ranked lists.
It compares the number of concordant pairs with the number of discordant pairs between each list.
A concordant pair is defined over two observations $(x_i, y_i)$ and $(x_j, y_j)$.
If $x_i > x_j$ and $y_i > y_j$, then the pair at indices $i,j$ is concordant, that is, the ranking at $i,j$ in both ranking sets $X$ and $Y$ agree with each other.
Similarly, a pair $i,j$ is discordant if $x_i > x_j$ and $y_i < y_j$ or $x_i < x_j$ and $y_i > y_j$.
Kendall's $\tau$ is then defined as:
\begin{equation}
  \tau = \dfrac{C - D}{\frac{1}{2}n(n-1)}
\end{equation}
where C is the number of concordant pairs, D is the number of discordant pairs, and the denominator represents the total number of possible pairs.
Thus, Kendall's $\tau$ falls in the range of $[-1,1]$, where -1 means that the ranked lists are perfectly negatively correlated, 0 means that they are not significantly correlated, and 1 means that the ranked lists are perfectly correlated.
One downside of this measure is that it does not take into account where in the ranked list an error occurs.
Information retrieval, in general, cares more about errors near the top of the list rather than errors near the bottom of the list.

\paragraph{Average-Precision $\tau$ ($\tau_{ap}$)}
Average-Precision $\tau$ (\cite{Yilmaz:2008bp}) (or $\tau_{ap}$) extends on Kendall's $\tau$ by incorporating the position of errors.
If an error occurs near the top of the list, then that is penalized heavier than an error occurring at the bottom of the list.
To achieve this, $\tau_{ap}$ incorporates ideas from the popular Average Precision measure, were we calculate the precision at each index of the list and then average them together.
$\tau_{ap}$ is defined as:
\begin{equation}
  \tau_{ap} = \dfrac{2}{N-1} \sum\limits_{i=2}^N \left(\dfrac{C(i)}{i-1}\right) - 1
\end{equation}
Intuitively, if an error occurs at the top of the list, then that error is propagated into each iteration of the summation, meaning that it's penalty is added multiple times.
$\tau_{ap}$'s range is between -1 and 1, where -1 means the lists are perfectly negatively correlated, 0 means that they are not significantly correlated, and 1 means that they are perfectly correlated.

\subsection*{Forward Feature Selection}
Forward feature selection was performed by iteratively testing each feature one at a time.
The highest performing feature is kept in the model, and another sweep is done over the remaining features.
This continues until all features have been selected.
This approach allows us to explore the effect of combinations of features and the effect of having too many or too few features.
It also allows us to evaluate which features and combinations of features are the most powerful.

\section*{Results}
We first compare our gold standard to the baselines.
A random baseline is provided for reference.
Because all of the documents that we rank are relevant, NDCG will be fairly high simply by chance.
We find that the number of times a document is mentioned in the annotated document is significantly better than the random baseline or the citation impact.
The more times a document is mentioned in a paper, the more likely the author was to annotate it as important.
Interestingly, we see a negative correlation with the citation impact.
The more times a document is mentioned in the literature, the less likely it is to be important.
These results are shown in Table~\ref{baseline-table}.


\begin{table*}[ht]
\centering
\begin{tabular}{*4l}
\toprule
{} &  NDCG & Kendall's $\tau$ & $\tau_{ap}$ \\
\midrule
Random Baseline   &  0.74 & -0.001   & 0.08\\
\# times referenced in document &	0.77 & 0.18 &	0.27\\
Citation Impact & 0.70 & -0.10 & -0.01\\
\bottomrule
\end{tabular}
\caption{Results for the citation baselines.
The number of times a citation is mentioned in the document is a better indicator of rank than the citation impact.
}
\label{baseline-table}
\end{table*}

Next we rank the raw values of the features and compare them to our gold standard to obtain a baseline (Table~\ref{feature-results-table}).
The best performing text similarity feature is the similarity between the abstract of the annotated document and the abstract of the cited document.
However, the number of times that a cited document is mentioned in the text of the annotated document are also high-scoring features, especially in the $\tau_{ap}$ correlation coefficient.
These results indicate that text similarity alone may not be a good measure for judging the rank of a document.

\begin{table*}[ht]
\centering
\begin{tabular}{*4l}
\toprule
Feature & NDCG & Kendall's $\tau$ & $\tau_{ap}$ \\
\midrule
Similarity(a,a) & 0.79 & 0.25 & 0.31\\
Similarity(a,t)  & 0.75 & 0.11 & 0.18\\
Similarity(t,a)  & 0.76 & 0.13 & 0.20\\
Similarity(t,t)  & 0.75 & 0.11 & 0.18\\
Similarity(f,a)  & 0.79 & 0.25 & 0.31\\
Similarity(f,t)  & 0.74 & 0.10 & 0.17\\
Similarity(c,a)  & 0.76 & 0.12 & 0.22\\
Similarity(c,t)  & 0.74 & 0.06 & 0.14\\
Similarity(d,a) & 0.78 & 0.20 & 0.27\\
Similarity(d,t) & 0.75 & 0.09 & 0.17\\
Similarity(cd,a)  & 0.76 & 0.12 & 0.20\\
Similarity(cd,t) & 0.75 & 0.07 & 0.15\\
Age & 0.69 & -0.12 & -0.012\\
MentionCount(f) & 0.77 & 0.18 & 0.27\\
MentionCount(d) & 0.76 & 0.12 & 0.21\\
CitationImpact & 0.70 & -0.10 & -0.013 \\
\bottomrule
\end{tabular}
\caption{Results for ranking by each individual feature value.
Similarity features are text similarity features.
The first parameter is the section of text in the annotated document, the second parameter is the section of text in the referenced document.
Here, ``a'' means abstract, ``t'' means title, ``f'' means full text, ``c'' means the 10 word window around a citation, ``d'' means the discussion/conclusion sections, and ``cd'' means 10 word windows around citations in the discussion/conclusion section.
Age is the age of the referenced document, MentionCount is the number of times the annotated document mentions the referenced document in text, and CitationImpact is the number of documents that have cited the referenced document in the literature.
}
\label{feature-results-table}
\end{table*}

Next we test three different feature sets for our supervised learning-to-rank models.
The model using only the text similarity features performs poorly: NDCG stays at baseline and the correlation measures are low.
Models that incorporate information about the age, number of times a cited document was referenced, and the citation impact of that document in addition to the text similarity features significantly outperformed models that used only text similarity features $\tau_{ap} = 0.35$.
Because $\tau_{ap}$ takes into account the position in the ranking of the errors, this indicates that the All Features model was able to better correctly place highly ranked documents above lower ranked ones.
Similarly, because Kendall's $\tau$ is an overall measure of correlation that does not take into account the position of errors, the higher value here means that more rankings were correctly placed.
Interestingly, feature selection (which is optimized for NDCG) does not outperform the model using all of the features in terms of our correlation measures.
The features chosen during forward feature selection are (1) the citation impact, (2) number of mentions in the full text, (3) text similarity between the annotated document's title and the referenced document's abstract, (4) the text similarity between the annotated document's discussion/conclusion section and the referenced document's title.
These results are shown in Table \ref{model-table}.
The models trained on the text similarity judgments perform worse than the models trained on the annotated data.
However, in terms of both NDCG and the correlation measures, they perform significantly better than the random baseline.

\begin{table*}[ht]
\centering
\begin{tabular}{*7l}
\toprule
{} &  \multicolumn{3}{c}{Annotated Model} & \multicolumn{3}{c}{Text Similarity Model}\\
\midrule
{} & NDCG & $\tau$ & $\tau_{ap}$ & NDCG & $\tau$ & $\tau_{ap}$ \\
\midrule
Text Features & 0.74 & 0.10 & 0.23 & 0.73 & 0.10 & 0.25\\
All Features* & 0.79 & \textbf{0.27} & \textbf{0.35} & 0.76 & 0.18 & 0.28\\
Feature Selection* & \textbf{0.80} & 0.26 & 0.33 & 0.78 & 0.20 & 0.30\\
\bottomrule
\end{tabular}
\caption{Results for the SVM\textsuperscript{Rank} models for three different combinations of features.
``Text Only Features'' are only the text similarity features.
``Feature Selection'' is the set of features found after running a forward feature selection algorithm.
A ``*'' indicates statistical significance between the two models.
}
\label{model-table}
\end{table*}

Next we compare our model to Google Scholar's rankings.
Using the ranking collected from Google Scholar, we build a training set to try to predict our authors' rankings.
We find that Google Scholar performs similarly to the text-only features model.
This indicates that the rankings we obtained from the authors are substantially different than the rankings that Google Scholar provides.
Results appear in Table \ref{scholar-table}.

\begin{table*}[t]
\centering
\begin{tabular}{*5l}
\toprule
Training Set & Test Set & NDCG & Kendall's $\tau$ & $\tau_{ap}$ \\
\midrule
Author Annotations & Author Annotations & 0.79 & 0.27 & 0.35\\
Google Scholar & Author Annotations & 0.77 & 0.16 & 0.27\\
\bottomrule
\end{tabular}
\caption{Results for the model trained using the Google Scholar Related Articles ranking.
We find that building a model using Google Scholar's Related Articles ranking to predict our authors' rankings performs poorly compared to the other models.}
\label{scholar-table}
\end{table*}

\section*{Discussion}
We found that authors rank the references they cite substantially differently from rankings based on text-similarity.
Our results show that decomposing a document into a set of features that is able to capture that difference is key.
While text similarity is indeed important (as evidenced by the Similarity(a,a) feature in Table \ref{feature-results-table}), we also found that the number of times a document is referenced in the text and the number of times a document is referenced in the literature are also both important features (via feature selection).
The more often a citation is mentioned in the text, the more likely it is to be important.
This feature is often overlooked in article citation recommendation.
We also found that recency is important: the age of the citation is negatively correlated with the rank.
Newer citations are more likely to be directly important than older, more foundational citations.
Additionally, the number of times a document is cited in the literature is negatively correlated with rank.
This is likely due to highly cited documents being more foundational works; they may be older papers that are important to the field but not directly influential to the new work.

The model trained using the author's judgments does significantly better than the model trained using the text-similarity-based judgments.
An error analysis was performed to find out why some of the rankings disagreed with the author's annotations.
We found that in some cases our features were unable to capture the relationship: for example a biomedical document applying a model developed in another field to the dataset may use very different language to describe the model than the citation.
Previous work adopting topic models to query document search may prove useful for such cases.

A small subset of features ended up performing as well as the full list of features.
The number of times a citation was mentioned and the citation impact score in the literature ended up being two of the most important features.
Indeed, without the citation-based features, the model performs as though it were trained with the text-similarity rankings.
Feature engineering is a part of any learning-to-rank system, especially in domain-specific contexts.
Citations are an integral feature of our dataset.
For learning-to-rank to be applied to other datasets feature engineering must also occur to exploit the unique properties of those datasets.
However, we show that combining the domain-specific features with more traditional text-based features does improve the model's scores over simply using the domain-specific features themselves.

Interestingly, citation impact and age of the citation are both negatively correlated with rank.
We hypothesize that this is because both measures can be indicators of recency: a new publication is more likely to be directly influenced by more recent work.
Many other related search tools, however, treat the citation impact as a positive feature of relatedness: documents with a higher citation impact appear higher on the list of related articles than those with lower citation impacts.
This may be the opposite of what the user actually desires.

We also found that rankings from our text-similarity based IR system or Google Scholar's IR system were unable to rank documents by the authors' annotations as well as our system.
In one sense, this is reasonable: the rankings coming from these systems were from a different system than the author annotations.
However, in domain-specific IR, domain experts are the best judges.
We built a system that exploits these expert judgments.
The text similarity and Google Scholar models were able to do this to some extent, performing above the random baseline, but not on the level of our model.

Additionally, we observe that NDCG may not be the most appropriate measure for comparing short ranked lists where all of the documents are relevant to some degree.
NDCG gives a lot of credit to relevant documents that occur in the highest ranks.
However, all of the documents here are relevant, just to varying degrees.
Thus, NDCG does not seem to be the most appropriate measure, as is evident in our scores.
The correlation coefficients from Kendall's $\tau$ and $\tau_{ap}$ seem to be far more appropriate for this case, as they are not concerned with relevance, only ranking.

One limitation of our work is that we selected a small set of references based on their similarities to the article that cites them.
Ideally, we would have had authors rank all of their citations for us, but this would have been a daunting task for authors to perform.
We chose to use the Google Scholar dataset in order to attempt to mitigate this: we obtain a ranking for the set of references from a system that is also ranking many other documents.
The five citations selected by TF*IDF weighted cosine similarity represent a ``hard'' gold standard: we are attempting to rank documents that are known to all be relevant by their nature, and have high similarity with the text.
Additionally, there are plethora of other, more expensive features we could explore to improve the model.
Citation network features, phrasal concepts, and topic models could all be used to help improve our results, at the cost of computational complexity.

We have developed a model for fast related-document ranking based on crowd-sourced data.
The model, data, and data collection software are all publicly available \footnote{http://github.com/umassbionlp/crowd-ranking.git} and can easily be used in future applications as an automatic search to help users find the most important citations given a particular document.
The experimental setup is portable to other datasets with some feature engineering.
We were able to identify that several domain-specific features were crucial to our model, and that we were able to improve on the results of simply using those features alone by adding more traditional features.

Query-by-document is a complicated and challenging task.
We provide an approach with an easily obtained dataset and a computationally inexpensive model.
By working with biomedical researchers we were able to build a system that ranks documents in a quantitatively different way than previous systems, and to provide a tool that helps researchers find related documents.

\section*{Acknowledgments}
We would like to thank all of the authors who took the time to answer our citation ranking survey.
This work is supported by National Institutes of Health with the grant number 1R01GM095476. The funders had no role in study design, data collection and analysis, decision to publish, or preparation of the manuscript.

\bibliography{LearningToRankSciDocs}

\end{document}